# Are volatility estimators robust with respect to modeling assumptions?

YINGYING LI[*] and PER A. MYKLAND[**]

*Department of Statistics, The University of Chicago, Chicago, IL 60637, USA.*
*E-mail: [*]yyli@galton.uchicago.edu; [**]mykland@galton.uchicago.edu*

We consider microstructure as an arbitrary contamination of the underlying latent securities price, through a Markov kernel $Q$. Special cases include additive error, rounding and combinations thereof. Our main result is that, subject to smoothness conditions, the two scales realized volatility is robust to the form of contamination $Q$. To push the limits of our result, we show what happens for some models that involve rounding (which is not, of course, smooth) and see in this situation how the robustness deteriorates with decreasing smoothness. Our conclusion is that under reasonable smoothness, one does not need to consider too closely how the microstructure is formed, while if severe non-smoothness is suspected, one needs to pay attention to the precise structure and also the use to which the estimator of volatility will be put.

## 1. Introduction

Recent years have seen an explosion of literature on the problem of estimating integrated volatility and similar objects with the help of high-frequency data. For a sample of recent literature, see Hull and White [16], Jacod and Protter [18], Gallant *et al.* [11], Chernov and Ghysels [7], Gloter [12], Andersen *et al.* [3], Dacorogna *et al.* [8], Barndorff-Nielsen and Shephard [6] and Mykland and Zhang [22], among others. An important realization has been that log prices do not appear to be semimartingales, but rather are like semimartingales observed with error. The main hypothesis proposed in the literature is that this error occurs by rounding (Delattre and Jacod [9]; Jacod [17]; Zeng [25]) or by additive error (Zhou [28]; Zhang *et al.* [27]; Zhang [26]; Aït-Sahalia *et al.* [1]; Bandi and Russell [5]; Hansen and Lunde [14]). More complex (and descriptive) models for microstructure are also available; see, for example, Hasbrouck [15] and, from a very different perspective, Farmer *et al.* [10].

The multiplicity of ways in which errors can be modeled raises the question of how sensitive inference is to modeling assumptions. This is the topic of this paper.

We shall be making the assumption that there is a latent log price process $X_t$ that is a continuous semimartingale of the form

$$\mathrm{d}X_t = \mu_t \, \mathrm{d}t + \sigma_t \, \mathrm{d}B_t, \tag{1}$$

where $\mu_t$ and $\sigma_t$ are continuous random processes, $\sigma_t$ is nonzero and $B_t$ is a Brownian motion. This is also called an Itô process. Transactions at times $0 = t_0 < t_1 < \cdots < t_n = T$ give rise to log prices $Y_{t_i}$ that are contaminated versions of $X_{t_i}$ as follows. We suppose that there is a family $Q(x, \mathrm{d}y)$ of conditional distributions so that, given $X_{t_i}$, the law of $Y_{t_i}$ is

$$P(Y_{t_i} \leq y \mid X \text{ process}) = P(Y_{t_i} \leq y \mid X_{t_i}) = Q(X_{t_i}, y). \tag{2}$$

In other words, $Y_{t_i}$ is distributed around $X_{t_i}$ in a way that depends only on the latter. We also assume that $Y_{t_0}, \ldots, Y_{t_n}$ are conditionally independent given the $X$ process.

A simple example of such contamination $Q$ is additive error on the log scale. If $Y = X + \varepsilon$, where $\varepsilon$ has density $g$ and is independent of $X$, then

$$Q(x, \mathrm{d}y) = g(y - x) \, \mathrm{d}y. \tag{3}$$

Another example is rounding or truncation. In this case, the probability distribution $Q(x, \mathrm{d}y)$ represents a non-random distortion of $x$. We shall look at yet another form of contamination in Section 3. In that case, the distortion is a combination of additive error and rounding.

This paper has two pieces of news: one good and one bad. We shall see that for reasonable types of contaminations $Q$, we can act as if the error is simply of additive type, and we shall see that the two scales realized volatility (TSRV) of Zhang *et al.* [27] is substantially robust to arbitrary contamination. This is our plan for Section 2.

There are, however, cases when we have to exercise care. We see one such case in Section 3, where we show that it is not always quite clear what is meant by volatility, and we have to consider carefully what quantity we actually wish to estimate. This occurs in cases that involve rounding.

We are mainly using TSRV as an example of a volatility estimator, and we believe that similar conclusions will apply, for example, to the multiscale realized volatility (MSRV) of Zhang [26]. With caveats about additional bias and variance, similar conclusions will also apply to traditional realized volatility (RV).

## 2. Robustness and smoothness of contamination

### 2.1. Setup

Suppose the latent log price process $X$ follows (1). Let $Y$ be the logarithm of the transaction price, which is observed at times $0 = t_0 < t_1 < \cdots < t_n = T$. We assume that at these sampling times, $Y$ is related to the latent log price process $X$ through (2). Let $f(X_t)$ be the conditional expectation of $Y_t$ given the $X$ process:

$$f(X_t) = \mathrm{E}_Q(Y_t | X_t). \tag{4}$$

We assume that

$$f(x) \text{ is twice continuously differentiable with } f'(x) \neq 0 \qquad \forall x. \tag{5}$$

**Definition 1.** *For two generic processes $Z^{(1)}$ and $Z^{(2)}$ and for an arbitrary grid $\mathcal{H} = \{s_0, s_1, s_2, \ldots, s_m\}$ of points in the interval $[0, T]$, define*

$$[Z^{(1)}, Z^{(2)}]_T^{\mathcal{H}} = \sum_{j=1}^{m} (Z_{s_j}^{(1)} - Z_{s_{j-1}}^{(1)})(Z_{s_j}^{(2)} - Z_{s_{j-1}}^{(2)}).$$

**Definition 2.** *If $Z$ is a continuous semimartingale, its quadratic variation $\langle Z, Z \rangle_T$ is defined as the limit in probability of $[Z, Z]_T^{\mathcal{H}_m}$ if $\mathcal{H}_m$ becomes dense in $[0, T]$ as $m \to \infty$. The quadratic variation is also known as the (integrated) volatility of $Z$ for the time period $[0, T]$.*

The above definition gives a well-defined limit $\langle Z, Z \rangle_T$ (independent of the sequence $\mathcal{H}_m$) in view of Theorems 4.47 and 4.48 of Jacod and Shiryaev ([19], 52).

A central problem is that we have two continuous semimartingale processes, $X$ and $f(X)$, that produce two volatilities:

$$\langle X, X \rangle_T = \int_0^T \sigma_t^2 \, \mathrm{d}t \quad \text{and} \quad \langle f(X), f(X) \rangle_T = \int_0^T f'(X_t)^2 \sigma_t^2 \, \mathrm{d}t$$

(see Protter [23], Theorem 29, pages 75–76). An interesting question arises immediately: which volatility are the volatility estimators estimating? When we make use of the observations $Y_{t_i}$ to estimate the volatility, we might think that we are estimating $\langle X, X \rangle_T$, because $Y_{t_i}$ is just the contaminated version of $X_{t_i}$, but given the $X$ process, $Y_{t_i}$ is centered at $f(X_{t_i})$, rather than $X_{t_i}$. We note that because both $X_t$ and $f(X_t)$ are Itô processes, without further model assumptions, we have nothing in the model that can answer the question of which volatility is the true underlying one.

These two volatilities $\langle X, X \rangle_T$ and $\langle f(X), f(X) \rangle_T$ are often similar quantities if $f(x) \approx x$, which makes the above question not so crucial, but this may not always be the case. Our first objective is to clarify which volatility the volatility estimators are estimating and how good the approximations are.

TSRV is a typical example of volatility estimators. For the moment, we focus on determining the properties of TSRV. We make use of some of the notations from Zhang *et al.* [27]:

Let $\mathcal{G} = \{0 = t_0, t_1, t_2, \ldots, t_n = T\}$ be the grid that contains all the observation times. We suppose $\mathcal{G}$ is partitioned into $K$ non-overlapping subgrids $\mathcal{G}^{(k)}$, $k = 1, \ldots, K$. As introduced in Zhang *et al.* [27], a typical example of selecting the subgrids is to use the regular allocation

$$\mathcal{G}^{(k)} = \{t_{k-1}, t_{k-1+K}, t_{k-1+2K}, \ldots, t_{k-1+n_k K}\}.$$

Let $n_k = |\mathcal{G}^{(k)}|$, the integer making $t_{k-1+n_k K}$ the last element in $\mathcal{G}^{(k)}$, and let $\bar{n} = \frac{1}{K} \sum_{k=1}^{K} n_k = \frac{1}{K}(n - K + 1)$. Further define $[Z^{(1)}, Z^{(2)}]_T^{\text{(all)}} = [Z^{(1)}, Z^{(2)}]_T^{\mathcal{G}}$ and $[Z^{(1)}, Z^{(2)}]_T^{\text{(avg)}} = \frac{1}{K} \sum_{k=1}^{K} [Z^{(1)}, Z^{(2)}]_T^{\mathcal{G}^{(k)}}$ for the two processes $Z^{(1)}$ and $Z^{(2)}$.

Estimators of the form $[Y, Y]_T^{\mathcal{H}}$ with $\mathcal{H} \subset \mathcal{G}$ are usually known as the RV. The TSRV is given by

$$\widehat{\langle X, X \rangle}_T = [Y, Y]_T^{\text{(avg)}} - \frac{\bar{n}}{n}[Y, Y]_T^{\text{(all)}}. \tag{6}$$

We assume constant step size ($\Delta t_i = T/n$) and that as $n \to \infty$,

$$K \to \infty \quad \text{and} \quad n/K \to \infty. \tag{7}$$

Note that our results generalize quite predictably if we allow $\Delta t_i$ to vary (see the theory in Zhang *et al.* [27]).

We also assume that the filtration for $(X_t)$ satisfies the "usual conditions" (see, e.g., Definition 1.3 of Jacod and Shiryaev [19], page 2) and Condition E of Zhang *et al.* [27].

## 2.2. Estimators of volatilities—estimators of $\langle f(X), f(X) \rangle_T$

Denote

$$\varepsilon_{t_i} = Y_{t_i} - f(X_{t_i}). \tag{8}$$

Note that under (2), the conditional moments of $\varepsilon_{t_i}$ depend only on the value of $X_{t_i}$. We assume that the conditional second moment of $\varepsilon_{t_i}$ is continuous and that there exists $\delta_0 > 0$ such that the conditional $(4 + 2\delta_0)$th moment of $\varepsilon_{t_i}$ is bounded on compact sets; that is,

$$g(x) := \mathrm{E}(\varepsilon_{t_i}^2 | X_{t_i} = x) \text{ is continuous,} \tag{9}$$

$$\forall l > 0, \exists M_{(4+2\delta_0, l)}, \text{ s.t. } \mathrm{E}(|\varepsilon_{t_i}|^{4+2\delta_0} | X_{t_i} = x) \leq M_{(4+2\delta_0, l)}, \text{ when } x \in [-l, l]. \tag{10}$$

We shall need the concept of stable convergence, as follows.

**Definition 3.** *Consider the $\sigma$-field $\Xi = \sigma(X_s, 0 \leq s \leq T)$. We say that a sequence $\zeta_n$ converges stably to $\zeta$ provided, for all $F \in \Xi$ and all bounded continuous $g$, $\mathrm{E} I_F g(\zeta_n) \to \mathrm{E} I_F g(\zeta)$ as $n \to \infty$, where $\zeta$ is defined on an extension of the original space.*

Note that because $X$ is continuous, the stable convergence of a sequence $\zeta_n$ is equivalent to the joint convergence of $\zeta_n$ with the process $X_s, 0 \le s \le T$ (see Jacod and Protter [18]; Section 2). This would not have been the case if $X$ were discontinuous. Also, note that we are using a specific reference $\sigma$-field $\Xi$, which is a little different from standard usage.

**Theorem 1.** *When we take $K = cn^{2/3}$ (the best possible order of TSRV), under the setup assumptions in Section 2.1, suppose (9) and (10) are satisfied, as $n \to \infty$,*

$$n^{1/6}(\langle \widehat{X, X} \rangle_T - \langle f(X), f(X) \rangle_T) \xrightarrow{\mathcal{L}} \left( \frac{8}{Tc^2} \int_0^T g(X_t)^2 \, \mathrm{d}t + c\xi^2 T \right)^{1/2} N(0, 1)$$

*stably, where*

$$\xi^2 = \frac{4}{3} \int_0^T (f'(X_t)\sigma_t)^4 \, \mathrm{d}t. \tag{11}$$

It is clear from this result what changes and what does not change for this more general contamination, compared to the case of independent additive error studied by Zhang *et al.* [27].

- The volatility being estimated is that of $f(X_t)$. (In Zhang *et al.* [27], $f(x) = x$.)
- The rate of convergence $n^{1/6}$ is the same as for independent additive error.
- The asymptotic variance changes to reflect the more complex form of contamination.

In summary, if we are happy to estimate the volatility of $f(X_t)$, the TSRV is exceedingly robust. The point about asymptotic variance is an issue only if we wish to set an interval around the observation. As can be seen from Zhang *et al.* [27], this is difficult even with straight additive contamination.

## 2.3. Proof of Theorem 1

We need to do some preparations before proving Theorem 1.

First, note that under the assumption (10), the following statements are true:

$$\forall \theta < 4 + 2\delta_0 \qquad \mathrm{E}(|\varepsilon_{t_i}|^\theta | X_{t_i} = x) \text{ is bounded on } [-l, l] \tag{12}$$

(we write the bound as $M_{(\theta, l)}$) and

$$\mathrm{Var}(\varepsilon_{t_i}^2 | X_{t_i} = x) = \mathrm{E}(\varepsilon_{t_i}^4 | X_{t_i} = x) - \mathrm{E}^2(\varepsilon_{t_i}^2 | X_{t_i} = x) \text{ is bounded on } [-l, l] \tag{13}$$

(say, by $M_{(\mathrm{Var}, l)}$). We shall use these notations in the proof:

$$M_T^{(1)} = \frac{1}{\sqrt{n}} \sum_{t_i \in \mathcal{G}} (\varepsilon_{t_i}^2 - \mathrm{E}(\varepsilon_{t_i}^2 | X));$$

$$M_T^{(2)} = \frac{1}{\sqrt{n}} \sum_{t_i \in \mathcal{G}} \varepsilon_{t_i} \varepsilon_{t_{i-1}};$$

$$M_T^{(3)} = \frac{1}{\sqrt{n}} \sum_{k=1}^{K} \sum_{t_i \in \mathcal{G}^{(k)}} \varepsilon_{t_i} \varepsilon_{t_{i,-}},$$

where $t_{i,-}$ denotes the previous element in $\mathcal{G}^{(k)}$ when $t_i \in \mathcal{G}^{(k)}$. $\varepsilon_{t_{-1}} = 0$, $\varepsilon_{t_{i,-}} = 0$ for $t_i = \min \mathcal{G}^{(k)}$.

**Proposition 1.** *Assume that* $\mathrm{E}(|A_n||X)$ *is* $O_P(1)$. *Then* $A_n$ *is* $O_P(1)$.

**Proof.** We have

$$P(|A_n| > K) \leq P(|A_n| I_{\{\mathrm{E}(|A_n||X) \leq K'\}} > K) + P(\mathrm{E}(|A_n||X) > K')$$

$$\leq \frac{\mathrm{E}(|A_n| I_{\{\mathrm{E}(|A_n||X) \leq K'\}})}{K} + P(\mathrm{E}(|A_n||X) > K')$$

$$= \frac{\mathrm{E}(\mathrm{E}(|A_n||X) I_{\{\mathrm{E}(|A_n||X) \leq K'\}})}{K} + P(\mathrm{E}(|A_n||X) > K')$$

$$\leq \frac{K'}{K} + P(\mathrm{E}(|A_n||X) > K')$$

for all $K, K'$. Hence the result follows. $\qquad\square$

**Lemma 1.** *We have*

$$[Y, Y]_T^{(\mathrm{all})} = [\varepsilon, \varepsilon]_T^{(\mathrm{all})} + O_P(1), \tag{14}$$

$$[Y, Y]_T^{(\mathrm{avg})} = [\varepsilon, \varepsilon]_T^{(\mathrm{avg})} + [f(X), f(X)]_T^{(\mathrm{avg})} + O_P\left(\frac{1}{\sqrt{K}}\right). \tag{15}$$

**Proof.** Define $\tau_l = \inf\{t : |X_t| \geq l\}$, $\forall l$. Note that $\tau_l$ has the property that

$$P(\tau_l \leq T) \to 0 \qquad \text{as } l \to \infty. \tag{16}$$

Also define $\Delta f(X_{t_i}) = f(X_{t_{i+1}}) - f(X_{t_i})$ for $i = 0, 1, \ldots, n-1$.

By (12), $g(X_t) = \mathrm{E}(\varepsilon_t^2 | X_t), t \leq T$ is bounded by $M_{(2,l)}$ on $\{\tau_l > T\}$; that is,

$$\mathrm{E}(([f(X), \varepsilon]_T^{(\mathrm{all})})^2 I_{\{\tau_l > T\}} | X)$$

$$= I_{\{\tau_l > T\}} \sum_{i=1}^{n-1} (\Delta f(X_{t_{i-1}}) - \Delta f(X_{t_i}))^2 \mathrm{E}(\varepsilon_{t_i}^2 | X) + \Delta f(X_{t_{n-1}})^2 \mathrm{E}(\varepsilon_{t_n}^2 | X)$$

$$+ \Delta f(X_{t_0})^2 \mathrm{E}(\varepsilon_{t_0}^2 | X)$$

$$\leq I_{\{\tau_l > T\}} M_{(2,l)} \left[ \sum_{i=1}^{n-1} (\Delta f(X_{t_{i-1}}) - \Delta f(X_{t_i}))^2 + \Delta f(X_{t_{n-1}})^2 + \Delta f(X_{t_0})^2 \right]$$

$$= 2 I_{\{\tau_l > T\}} M_{(2,l)} \left( [f(X), f(X)]_T^{(\mathrm{all})} - \sum_{i=1}^{n-1} \Delta f(X_{t_{i-1}}) \Delta f(X_{t_i}) \right)$$

$$\leq 4 I_{\{\tau_l > T\}} M_{(2,l)} [f(X), f(X)]_T^{(\mathrm{all})}$$

$$= O_P(1),$$

where we have used the Cauchy–Schwarz inequality. Hence, by Proposition 1 and (16),

$$[f(X), \varepsilon]_T^{(\mathrm{all})} = O_P(1). \tag{17}$$

Parallel argument shows that $\mathrm{E}(([f(X), \varepsilon]_T^{(\mathrm{avg})})^2 I_{\{\tau_l > T\}} | X) = O_p(\frac{1}{K})$. Hence,

$$[f(X), \varepsilon]_T^{(\mathrm{avg})} = O_P\left( \frac{1}{\sqrt{K}} \right). \tag{18}$$

Equalities (17) and (18) imply (14) and (15) because

$$[Y, Y]_T^{(\mathrm{all})} = [\varepsilon, \varepsilon]_T^{(\mathrm{all})} + [f(X), f(X)]_T^{(\mathrm{all})} + 2[f(X), \varepsilon]_T^{(\mathrm{all})}$$

and

$$[Y, Y]_T^{(\mathrm{avg})} = [f(X), f(X)]_T^{(\mathrm{avg})} + [\varepsilon, \varepsilon]_T^{(\mathrm{avg})} + 2[f(X), \varepsilon]_T^{(\mathrm{avg})}. \qquad \square$$

**Lemma 2.** $M_T^{(2)}$ *and* $M_T^{(3)}$ *are asymptotically independently normal conditionally on* $X$, *both with variance* $\frac{1}{T} \int_0^T g(X_t)^2 \, \mathrm{d}t$.

**Proof.** We use $\langle \cdot, \cdot \rangle_T$ to denote the discrete-time predictable quadratic variations and covariations (see Hall and Heyde [13], page 51) in this proof. Note that they are different from the continuous time quadratic variations in Definition 2. $\qquad \square$

$M_T^{(2)}$ and $M_T^{(3)}$ are the end-points of martingales with respect to filtration $\mathcal{F}_i = \sigma(\varepsilon_{t_j}, j \leq i, X_t, \text{all } t)$:

$$\langle M^{(2)}, M^{(2)} \rangle_T = \frac{1}{n} \sum_{t_i \in \mathcal{G}} \mathrm{Var}(\varepsilon_{t_i} \varepsilon_{t_{i-1}} | \mathcal{F}_{i-1})$$

$$= \frac{1}{n} \sum_{t_i \in \mathcal{G}} \varepsilon_{t_{i-1}}^2 g(X_{t_i})$$

$$= \frac{1}{n} \sum_{t_i \in \mathcal{G}} (\varepsilon_{t_{i-1}}^2 - g(X_{t_{i-1}})) g(X_{t_i}) + \frac{1}{T} \sum_{t_i \in \mathcal{G}} g(X_{t_{i-1}}) g(X_{t_i}) \Delta t. \tag{19}$$

Note that

$$\mathrm{E}\left(\left(\frac{1}{n}\sum_{t_i\in\mathcal{G}}(\varepsilon_{t_{i-1}}^2-g(X_{t_{i-1}}))g(X_{t_i})I_{\{\tau_l>T\}}\right)^2\Big|X\right)$$

$$=\mathrm{Var}\left(\frac{1}{n}\sum_{t_i\in\mathcal{G}}(\varepsilon_{t_{i-1}}^2-g(X_{t_{i-1}}))g(X_{t_i})I_{\{\tau_l>T\}}|X\right)$$

$$=\frac{1}{n^2}\sum_{t_i\in\mathcal{G}}\mathrm{Var}(\varepsilon_{t_{i-1}}^2|X)g^2(X_{t_i})I_{\{\tau_l>T\}}$$

$$\leq\frac{1}{n^2}\sum_{t_i\in\mathcal{G}}M_{(\mathrm{Var},l)}M_{(2,l)}$$

$$=O_P\left(\frac{1}{n}\right);$$

hence, by Proposition 1 and (16), the first term of (19) $\frac{1}{n}\sum_{t_i\in\mathcal{G}}(\varepsilon_{t_{i-1}}^2-g(X_{t_{i-1}}))g(X_{t_i})\to_P$ 0. Therefore, by (9),

$$\langle M^{(2)},M^{(2)}\rangle_T=\frac{1}{T}\sum_{t_i\in\mathcal{G}}g(X_{t_{i-1}})g(X_{t_i})\Delta t+o_p(1)\to_P\frac{1}{T}\int_0^T g(X_t)^2\,\mathrm{d}t.$$

Parallel argument shows that

$$\langle M^{(3)},M^{(3)}\rangle_T=\frac{1}{n}\sum_{k=1}^K\sum_{t_i\in\mathcal{G}^{(k)}}\mathrm{Var}(\varepsilon_{t_i}\varepsilon_{t_{i,-}}|\mathcal{F}_{i-1})\to_P\frac{1}{T}\int_0^T g(X_t)^2\,\mathrm{d}t.$$

On the other hand,

$$\langle M^{(2)},M^{(3)}\rangle_T=\frac{1}{n}\sum_{k=1}^K\sum_{t_i\in\mathcal{G}^{(k)}}\mathrm{Cov}(\varepsilon_{t_i}\varepsilon_{t_{i-1}},\varepsilon_{t_i}\varepsilon_{t_{i,-}}|\mathcal{F}_{i-1})=\frac{1}{n}\sum_{k=1}^K\sum_{t_i\in\mathcal{G}^{(k)}}\varepsilon_{t_{i-1}}\varepsilon_{t_{i,-}}\mathrm{E}(\varepsilon_{t_i}^2|X).$$

As a consequence,

$$\mathrm{E}((\langle M^{(2)},M^{(3)}\rangle_T)^2 I_{\{\tau_l>T\}}|X)=\frac{1}{n^2}\sum_{k=1}^K\sum_{t_i\in\mathcal{G}^{(k)}}\mathrm{E}^2(\varepsilon_{t_i}^2|X)\mathrm{E}(\varepsilon_{t_{i-1}}^2\varepsilon_{t_{i,-}}^2 I_{\{\tau_l>T\}}|X)$$

$$\leq\frac{1}{n^2}\sum_{k=1}^K\sum_{t_i\in\mathcal{G}^{(k)}}\mathrm{E}^2(\varepsilon_{t_i}^2|X)\sqrt{\mathrm{E}(\varepsilon_{t_{i-1}}^4 I_{\{\tau_l>T\}}|X)\mathrm{E}(\varepsilon_{t_{i,-}}^4 I_{\{\tau_l>T\}}|X)}$$

$$\leq\frac{1}{n}M_{(2,l)}^2 M_{(4,l)}$$

$$= O_P\left(\frac{1}{n}\right).$$

By Proposition 1 and (16), $\langle M^{(2)}, M^{(3)}\rangle_T \to_P 0$.

By assumption (10) and Proposition 1, we can easily see that the conditional Lyapunov conditions are satisfied. Also note that the limiting predictable quadratic variation $\frac{1}{T}\int_0^T g(X_t)^2\,dt$ and the limiting predictable quadratic covariation (which is zero) are measurable in (the completions of) all the $\sigma$-fields $\mathcal{F}_i$, so we can make use of the Remarks immediately following Corollary 3.1 (the martingale central limit theorem) in Hall and Heyde [13] to obtain the conclusion.

**Proof of Theorem 1.** Denote

$$R_1 = (\varepsilon_{t_0}^2 - \mathrm{E}(\varepsilon_{t_0}^2|X)) + (\varepsilon_{t_n}^2 - \mathrm{E}(\varepsilon_{t_n}^2|X))$$

and

$$R_2 = \sum_{k=1}^K [(\varepsilon_{\min \mathcal{G}^{(k)}}^2 - \mathrm{E}(\varepsilon_{\min \mathcal{G}^{(k)}}^2|X)) + (\varepsilon_{\max \mathcal{G}^{(k)}}^2 - \mathrm{E}(\varepsilon_{\max \mathcal{G}^{(k)}}^2|X))].$$

By (13), $\mathrm{E}(R_1^2 I_{\{\tau_1 > T\}}|X) = O_p(1)$; hence, $R_1 = O_p(1)$ by Proposition 1 and (16). Similarly, $\mathrm{E}(R_2^2 I_{\tau_1 > T}|X) = O_p(K)$; hence, $R_2 = O_p(K^{1/2})$. As a consequence,

$$[\varepsilon, \varepsilon]_T^{(\mathrm{all})} = 2\sum_{t_i \in \mathcal{G}} (\varepsilon_{t_i}^2 - \mathrm{E}(\varepsilon_{t_i}^2|X)) - 2\sum_{t_i > 0} \varepsilon_{t_i}\varepsilon_{t_{i-1}} + 2\sum_{t_i \in \mathcal{G}} E(\varepsilon_{t_i}^2|X) - R_1$$

$$= 2\sqrt{n}(M^{(1)} - M^{(2)}) + 2\sum_{t_i \in \mathcal{G}} \mathrm{E}(\varepsilon_{t_i}^2|X) + O_P(1)$$

and

$$K[\varepsilon, \varepsilon]_T^{(\mathrm{avg})} = 2\sqrt{n}(M^{(1)} - M^{(3)}) - R_2 + 2\sum_{k=1}^K \sum_{t_i \in \mathcal{G}^{(k)}} \mathrm{E}(\varepsilon_{t_i}^2|X)$$

$$= 2\sqrt{n}(M^{(1)} - M^{(3)}) + O_P(K^{1/2}) + 2\sum_{k=1}^K \sum_{t_i \in \mathcal{G}^{(k)}} \mathrm{E}(\varepsilon_{t_i}^2|X).$$

Therefore, conditionally on the $X$ process,

$$\frac{K}{\sqrt{n}}\left([\varepsilon, \varepsilon]_T^{(\mathrm{avg})} - \frac{\bar{n}}{n}[\varepsilon, \varepsilon]_T^{(\mathrm{all})}\right) \approx \frac{1}{\sqrt{n}}(K[\varepsilon, \varepsilon]_T^{(\mathrm{avg})} - [\varepsilon, \varepsilon]_T^{(\mathrm{all})})$$

$$= (2(M^{(2)} - M^{(3)})) + O_P\left(\sqrt{\frac{K}{n}}\right)$$

$$\xrightarrow{\mathcal{L}} N\left(0, \frac{8}{T}\int_0^T g(X_t)^2\,dt\right). \qquad (20)$$

Observe that

$$\langle \widehat{X, X} \rangle_T = [Y, Y]_T^{(\text{avg})} - \frac{\bar{n}}{n} [Y, Y]_T^{(\text{all})}$$

$$= [f(X), f(X)]_T^{(\text{avg})} + [\varepsilon, \varepsilon]_T^{(\text{avg})} + O_p\left(\frac{1}{\sqrt{K}}\right)$$

$$- \frac{\bar{n}}{n} [\varepsilon, \varepsilon]_T^{(\text{all})} - O_p\left(\frac{\bar{n}}{n}\right) \qquad \text{(by Lemma 1)}$$

$$= [f(X), f(X)]_T^{(\text{avg})} + [\varepsilon, \varepsilon]_T^{(\text{avg})} - \frac{\bar{n}}{n} [\varepsilon, \varepsilon]_T^{(\text{all})} + O_p\left(\frac{1}{\sqrt{K}}\right)$$

and note that $\frac{K}{\sqrt{n}} \sim \frac{K}{\sqrt{K\bar{n}}} \sim \sqrt{\frac{K}{\bar{n}}}$ by (20), conditionally on the $X$ process

$$\sqrt{\frac{K}{\bar{n}}} (\langle \widehat{X, X} \rangle_T - [f(X), f(X)]_T^{(\text{avg})}) \xrightarrow{\mathcal{L}} N\left(0, \frac{8}{T} \int_0^T g(X_t)^2 \, dt\right). \tag{21}$$

On the other hand, $f(X_t)$ is a semimartingale, $df(X_t) = (f'(X_t)\mu_t + \frac{1}{2} f''(X_t)\sigma_t^2) \, dt + f'(X_t)\sigma_t \, dB_t$. By Zhang *et al.* [27],

$$\sqrt{\frac{n}{K}} ([f(X), f(X)]_T^{(\text{avg})} - \langle f(X), f(X) \rangle_T) \xrightarrow{\mathcal{L}} \xi \sqrt{T} \cdot Z_{\text{discrete}}, \tag{22}$$

where $\xi$ is defined as in (11) and $Z_{\text{discrete}} \sim N(0, 1)$ is independent of the process $X$. The convergence in law is stable.

Combining (21) and (22), we have

$$\langle \widehat{X, X} \rangle_T - \langle f(X), f(X) \rangle_T = (\langle \widehat{X, X} \rangle_T - [f(X), f(X)]_T^{(\text{avg})})$$

$$+ ([f(X), f(X)]_T^{(\text{avg})} - \langle f(X), f(X) \rangle_T)$$

$$= O_p\left(\frac{\bar{n}^{1/2}}{K^{1/2}}\right) + O_p(\bar{n}^{-1/2}).$$

The error is minimized when $K = O(n^{2/3})$. If we take $K = cn^{2/3}$, we have [by exploiting the conditional convergence in (21)]

$$n^{1/6}(\langle \widehat{X, X} \rangle_T - \langle f(X), f(X) \rangle_T) \to_{\mathcal{L}} \left(\frac{8}{Tc^2} \int_0^T g(X_t)^2 \, dt + c\xi^2 T\right)^{1/2} N(0, 1) \text{ stably.} \quad \square$$

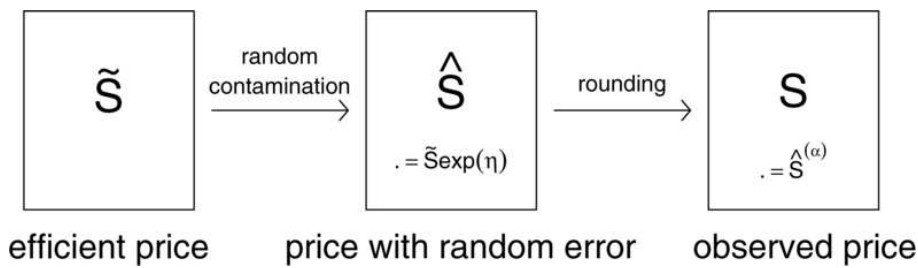

**Figure 1.** Two stage contamination: random error followed by rounding.

# 3. A case study

## 3.1. Another form of contamination

Two types of errors—additive errors and rounding errors—have been proposed as candidates of market microstructure errors. Results of when one of them plays the role are available. Now, we consider the case when both types of errors are present.

Suppose at the transaction times, the latent return process $X_{t_i}$ is contaminated by an independent additive error process $\eta_{t_i}$ and then rounded to reflect that prices are quotes on a grid (typically in multiples of one cent). We thus envision a two stage procedure where a latent efficient price $\tilde{S} = \exp(X)$ is first subjected to multiplicative random error: $\hat{S} = \tilde{S}\exp(\eta)$. The actual price $S$ is then the rounded value of $\hat{S}$. If we take, as usual, $Y = \log S$, our final product is the observed process $Y_{t_i}$ of logarithms of rounded contaminated prices

$$Y_{t_i} = \log((\exp(X_{t_i} + \eta_{t_i}))^{(\alpha)}),$$

where $s^{(\alpha)} = \alpha[s/\alpha]$ is the value of $s$ rounded to the nearest multiple of $\alpha$. The model is somewhat similar to that used by Large [21]. It can be illustrated as in Figure 1.

For practical purposes, we further assume that the smallest observation of the security price is $\alpha$, which makes observations of the log prices have the form

$$Y_{t_i} = \log\alpha \vee \log((\exp(X_{t_i} + \eta_{t_i}))^{(\alpha)}). \tag{23}$$

We consider the case when the random errors are independent identically distributed normal random variables, with mean 0 and positive variance; that is,

$$\eta_{t_i} \sim_{\text{i.i.d.}} N(0, \gamma^2), \qquad \gamma > 0. \tag{24}$$

In this case,

$$\begin{aligned}
f(x) &= \mathrm{E}(Y_{t_i} | X_{t_i} = x) \\
&= \int_{-\infty}^{\infty} \frac{1}{\sqrt{2\pi}\gamma}(\log\alpha \vee \log((e^z)^{(\alpha)}))e^{-(z-x)^2/(2\gamma^2)} \, \mathrm{d}z
\end{aligned} \tag{25}$$

is a twice continuously differentiable function with positive first derivative, and the assumptions (9) and (10) hold. Therefore, by Theorem 1, the TSRV is a robust estimator of $\langle f(X), f(X) \rangle_T$.

In this study we assume that $\alpha$ is a fixed quantity that is independent of the number of observations. Note that in the case where $\alpha \to 0$, we can expect relatively well posed behavior in view of Kolassa and McCullagh [20] and Delattre and Jacod [9].

## 3.2. Robustness works: When $\gamma$ is big

Assume that the latent price process $\tilde{S} = \exp(X_t)$ has a small probability of going below $\alpha$ for $t \in [0, T]$. Then, under model (23) and assumption (24), we have

$$f(X_t) \approx X_t \quad \text{and} \quad f'(X_t) \approx 1 \text{ for } t \in [0, T], \qquad \text{for suitably big } \gamma \text{s.}$$

By 'suitably big $\gamma$s,' we mean that the size of the random error is large enough that the possibility that it pulls the observations of the prices up or down several grid points (multiples of $\alpha$) is not negligible. In this case, when taking the conditional expectation, the positive and negative errors cancel out, and this leads to the result that $f(X_t) \approx X_t$.

In this case, $\langle f(X), f(X) \rangle_T \approx \langle X, X \rangle_T$. Therefore, the TSRV, which is a robust estimator of $\langle f(X), f(X) \rangle_T$, is a good estimator of $\langle X, X \rangle_T$ as well.

These relationships are illustrated in Section 3.4.

## 3.3. How things can go wrong: When $\gamma \to 0$

When $\gamma$ is small but not 0, by Theorem 1, we know that the TSRV goes to the limit $\langle f(X), f(X) \rangle_T$ robustly. However, this volatility $\langle f(X), f(X) \rangle_T$ is no longer close to $\langle X, X \rangle_T$.

To study the limiting behavior of $\langle f(X), f(X) \rangle_T$, we relate it to the local time $L_T^a$ of the semimartingale $X$ (for a definition of the local time, see Revuz and Yor [24], page 222). By Corollary 1.6 of Revuz and Yor ([24], page 224), for any positive Borel function $\Phi$, almost surely,

$$\int_0^t \Phi(X_s) \, d\langle X, X \rangle_s = \int_{-\infty}^{\infty} \Phi(a) L_t^a \, da. \tag{26}$$

By Exercise (1.32) of Revuz and Yor ([24], page 237), for the process $X$ given by (1), the family $L^a$ may be chosen such that

$$L_t^a \text{ is continuous in } a \text{ almost surely.} \tag{27}$$

We shall consider only the version of the local time $L_t^a$ that satisfies the condition (27).

Relating $\langle f(X), f(X) \rangle_T$ to $L_T^a$ by (26), we obtain the following result:

**Theorem 2.** *As $\gamma \to 0$, for the process $X$ given by (1) and $f$ defined as in (25), almost surely,*

$$\gamma \langle f(X), f(X) \rangle_T \to \frac{1}{2\sqrt{\pi}} \sum_{k=1}^{\infty} L_T^{\log((k+1/2)\alpha)} \left( \log \frac{k+1}{k} \right)^2,$$

*where $L_t^a$ is the local time of the continuous semimartingale $X$.*

In other words, the 'target' $\langle f(X), f(X) \rangle_T$ that we are estimating blows up as $\gamma$ goes to zero and is of order $1/\gamma$. This fact raises questions of whether $\langle f(X), f(X) \rangle_T$ is, in this case, the quantity that we are really seeking.

### 3.4. Illustration

We take a typical sample path to illustrate the situation: suppose the latent log price process $X_t$ follows (1) with $\mu_t = 0$ and $\sigma_t = 0.2, \forall t \in [0, \infty)$ (these are the annualized parameters). Suppose at observation time $t_i$, the price $\exp(X_{t_i})$ first is contaminated by an independent multiplicative random error $\exp(\eta_{t_i})$ with $\eta_{t_i}$ independent identically distributed as $N(0, \gamma^2)$ and then is rounded to the nearest multiple of $\alpha = 0.01$ (one cent). The quantity of interest is the volatility of the process over time $t \in [0, T]$ with $T = 1/252$ (i.e., one day). We assume that a day consists of 6.5 hours of open trading and that the price process is observed once every second ($n = 23\,400$).

A sample path of the latent log price process $X_t, t \in [0, T]$, is plotted in Figure 2, together with its corresponding pure rounded process (the solid line) and two $f(X_t)$ processes [see (25)] with $\gamma = 0.001$ (asterisks) and $\gamma = 0.005$ (open circles), respectively.

Figure 3 records the TSRV of this particular sample path $X_t, t \in [0, T]$, in Figure 2, with random contaminations of different sizes (with standard error $\gamma$ ranges from 0.0002 to 0.006). The solid line is the volatility $\langle X, X \rangle_T$.

We see from Figure 2 that for this process, when $\gamma$ is as large as 0.005, the $f(X)$ is close to $X$, while when $\gamma$ is smaller, the process $f(X)$ diverges from $X$; in fact, it goes closer to the (discontinuous) pure rounded process. Figure 3 shows that when $\gamma$ is suitably large, the TSRV can be a good estimator of $\langle X, X \rangle_T$, but when $\gamma$ is too small, the estimator does not estimate $\langle X, X \rangle_T$, but rather a much larger quantity. Note that although it is similarly shaped, this graph is not a signature plot in the sense of Andersen *et al.* [2], because the horizontal axis represents $\gamma$ rather than sample size. There is, however, a connection between these two types of plots, as shown in equation (29) below.

### 3.5. How error and sample size relate to each other—comparison to the case when $\gamma = 0$

When $\gamma = 0$, the additive error is gone, only the rounding error is present. In this case, the observations are themselves the conditional expectations and $f(x)$ is no longer continuous:

$$Y_{t_i} = f(X_{t_i}) = \mathrm{E}(Y_{t_i} | X_{t_i}) = \log \alpha \vee \log((\exp(X_{t_i}))^{(\alpha)}). \tag{28}$$

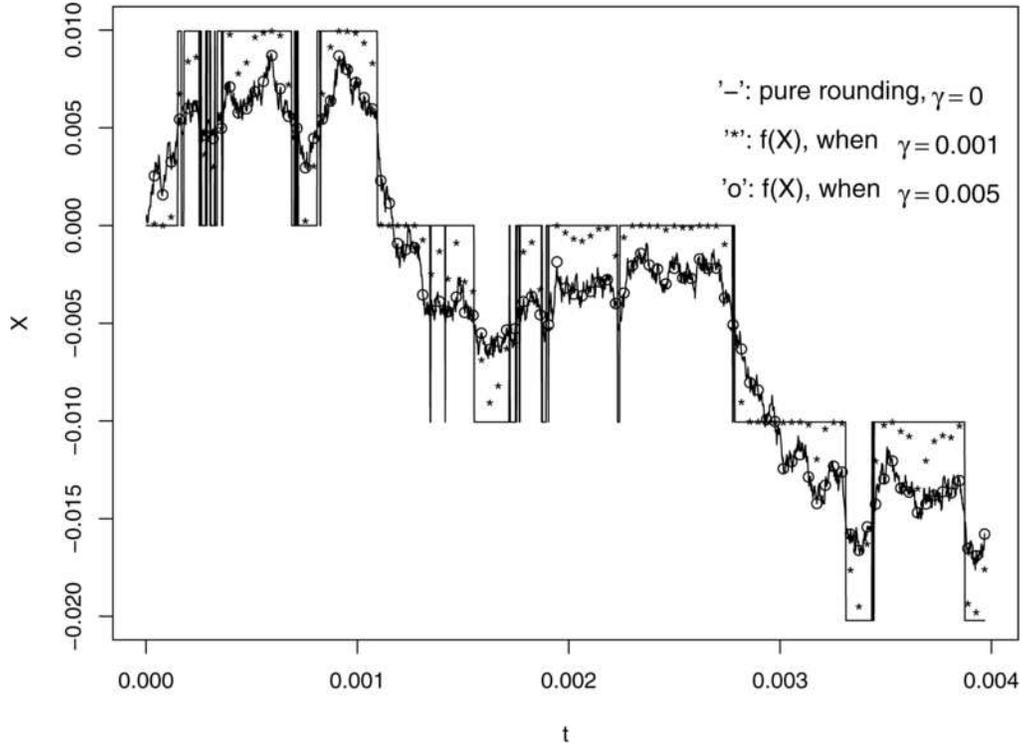

**Figure 2.** Relationship between $f(X)$ and $X$ on one (random) sample path.

Recall the notation from Section 2.1 about the TSRV $\widehat{\langle X, X \rangle}_T$. In particular, $K$ is the number of subgrids and $\bar{n} = \frac{1}{K}(n - K + 1)$ is the average number of elements in the subgrids. Also recall assumption (7), which is equivalent to '$\bar{n} \to \infty$ and $\bar{n}/n \to 0$ as $n \to \infty$.' A modification of the Jacod [17] proof gives the following result:

**Theorem 3.** *When $\gamma = 0$, if $X = \sigma W$ where $\sigma > 0$ and $W$ is a standard Brownian motion, we have*

$$\plim_{n \to \infty} \frac{1}{\sqrt{\bar{n}}} \widehat{\langle X, X \rangle}_T = \frac{1}{\sigma \sqrt{T}} \sqrt{\frac{2}{\pi}} \sum_{k=1}^{\infty} L_T^{\log((k+1/2)\alpha)} \left( \log \frac{k+1}{k} \right)^2,$$

*where $L_t^a$ is as in Theorem 2.*

We can see from Theorems 2 and 3 that to first order,

TSRV under pure rounding and no contamination

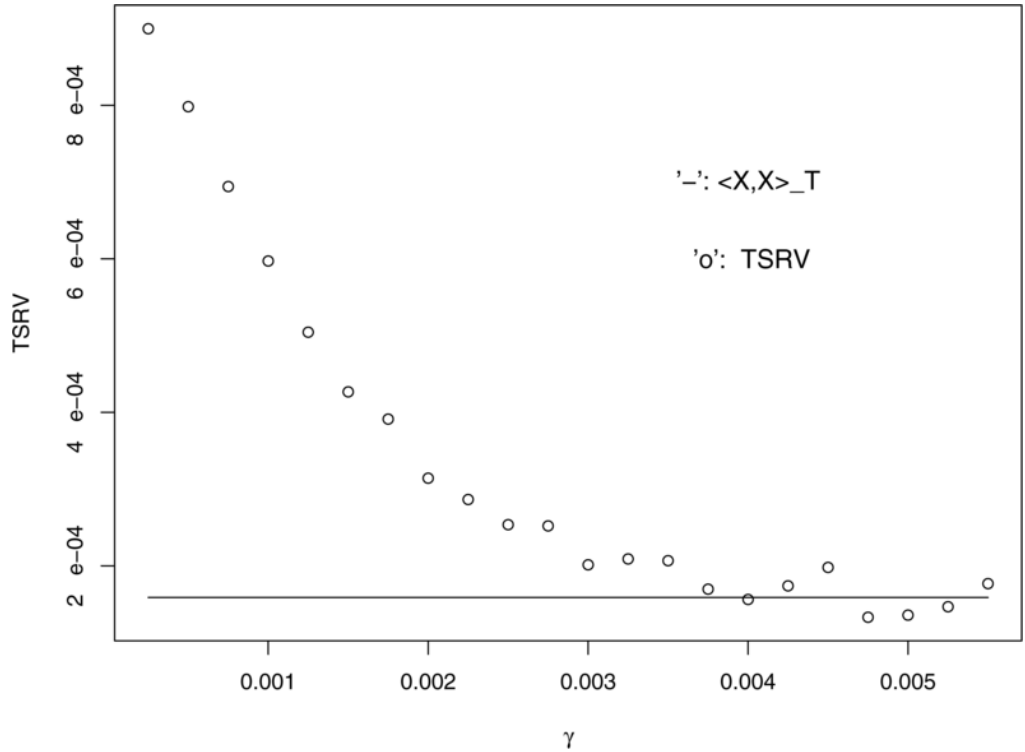

**Figure 3.** TSRV *vs* size of the random contamination, based on one (random) latent log price process $(\langle X, X \rangle_T = (0.2)^2 / 252 \approx 1.59 \cdot 10^{-4})$. For more details of the simulation, please refer to Section 3.4.

$$= \sqrt{\frac{8\bar{n}\gamma^2}{\sigma^2 T}} \times \text{TSRV under rounding after contamination of size } \gamma. \quad (29)$$

Thus, in a sense, contamination plays a role slightly similar to sample size when there is no contamination:

$$\gamma^{-2} \text{ under random contamination} \approx \bar{n}\frac{8}{\sigma^2 T} \text{ under no random contamination.} \quad (30)$$

In both cases, the sizes of of $\gamma^{-2}$ and $\bar{n}$ have similar functions in quantifying the ill-posedness of the respective estimation problems. The deeper meaning of this remains, for the moment, a little mysterious, even to us.

### 3.6. Proofs of Theorem 2 and Theorem 3

**Proof of Theorem 2.** We have, by (26), for $\mu(\mathrm{d}a) = \mu_\gamma(\mathrm{d}a) = (f'(a))^2 \, \mathrm{d}a$, almost surely,

$$\langle f(X), f(X) \rangle_T = \int_0^T (f'(X_t))^2 \, \mathrm{d}\langle X, X \rangle_t = \int_{-\infty}^\infty L_T^a \mu(\mathrm{d}a).$$

Recall that $f(x) = \int_{-\infty}^\infty \frac{1}{\sqrt{2\pi}\gamma} (\log \alpha \vee \log((\mathrm{e}^z)^{(\alpha)})) \mathrm{e}^{-((z-x)^2/(2\gamma^2))} \, \mathrm{d}z$; hence,

$$f'(x) = \int_{-\infty}^\infty \frac{1}{\sqrt{2\pi}\gamma} (\log \alpha \vee \log((\mathrm{e}^z)^{(\alpha)})) \frac{z-x}{\gamma^2} \mathrm{e}^{-((z-x)^2/(2\gamma^2))} \, \mathrm{d}z$$

$$= \int_{-\infty}^\infty \frac{1}{\gamma} \log \alpha \vee \log(\exp(x + \gamma v)^{(\alpha)}) \frac{v}{\sqrt{2\pi}} \mathrm{e}^{-v^2/2} \, \mathrm{d}v.$$

For $k = 1, 2, 3, \ldots$, we have, $\forall y \in \mathbb{R}$,

$$\gamma f'\left( \log\left( \left( k + \frac{1}{2} \right)\alpha \right) + y\gamma \right)$$

$$= \int_{-\infty}^\infty \log \alpha \vee \log\left( \exp\left( \log\left( \left( k + \frac{1}{2} \right)\alpha \right) + y\gamma + v\gamma \right)^{(\alpha)} \right) \frac{v}{\sqrt{2\pi}} \mathrm{e}^{-v^2/2} \, \mathrm{d}v$$

$$= \mathrm{E}_V\left( \log \alpha \vee \log\left( \left( \left( k + \frac{1}{2} \right)\alpha \cdot \mathrm{e}^{\gamma(y+V)} \right)^{(\alpha)} \right) \cdot V \right), \qquad V \sim N(0,1).$$

By the dominated convergence theorem,

$$\lim_{\gamma \to 0} \gamma f'\left( \log\left( \left( k + \frac{1}{2} \right)\alpha \right) + y\gamma \right)$$

$$= \mathrm{E}_V(\log((k+1)\alpha) \cdot VI_{\{y+V>0\}}) + \mathrm{E}_V(\log(k\alpha) \cdot VI_{\{y+V<0\}})$$

$$= \log((k+1)\alpha) \frac{1}{\sqrt{2\pi}} \int_{-y}^\infty z\mathrm{e}^{-z^2/2} \, \mathrm{d}z + \log(k\alpha) \frac{1}{\sqrt{2\pi}} \int_{-\infty}^{-y} z\mathrm{e}^{-z^2/2} \, \mathrm{d}z$$

$$= \frac{1}{\sqrt{2\pi}} \mathrm{e}^{-y^2/2} \log\left( \frac{k+1}{k} \right). \tag{31}$$

For $k = 1, 2, 3, \ldots$, denote $x_k = \log((k + \frac{1}{2})\alpha)$. For any $\delta_0 \in (0, \log(k + \frac{1}{2})/(k - \frac{1}{2}))$ and $\delta_1 \in (0, \log(k + \frac{3}{2})/(k + \frac{1}{2}))$,

$$\gamma\mu[x_k - \delta_0, x_k + \delta_1] = \int_{x_k - \delta_0}^{x_k + \delta_1} \gamma(f'(x))^2 \, \mathrm{d}x$$

$$= \int_{-\infty}^\infty (\gamma f'(x_k + y\gamma))^2 I_{\{y \in [-\delta_0/\gamma, \delta_1/\gamma]\}} \, \mathrm{d}y.$$

By the dominated convergence theorem,

$$
\lim_{\gamma \to 0} \gamma \mu[x_k - \delta_0, x_k + \delta_1] = \int_{-\infty}^{\infty} \lim_{\gamma \to 0} (\gamma f'(x_k + y\gamma))^2 I_{\{y \in [-\delta_0/\gamma, \delta_1/\gamma]\}} \, \mathrm{d}y
$$

$$
= \int_{-\infty}^{\infty} \left( \frac{1}{\sqrt{2\pi}} \log \frac{k+1}{k} \right)^2 \mathrm{e}^{-y^2} \, \mathrm{d}y
$$

$$
= \frac{1}{2\sqrt{\pi}} \left( \log \frac{k+1}{k} \right)^2. \tag{32}
$$

As a consequence, for any $\delta_0', \delta_1' \in (0, \frac{1}{2} \log{(k + \frac{3}{2})}/(k + \frac{1}{2}))$,

$$
\lim_{\gamma \to 0} \gamma \mu[x_k + \delta_0', x_{k+1} - \delta_1'] = 0. \tag{33}
$$

A simpler argument shows that for any $a < x_1$ and $\delta \in (0, x_1 - a)$,

$$
\lim_{\gamma \to 0} \gamma \mu[a, x_1 - \delta] = 0. \tag{34}
$$

Define $\nu$ to be the finite measure on $\mathbb{R}$ that has point mass $\frac{1}{2\sqrt{\pi}}(\log \frac{k+1}{k})^2$ on $x_k$, $k = 1, 2, 3, \ldots$. For any continuous function $\phi$ that vanishes outside a compact set, suppose that the support of $\phi$ is in $[-C, C]$ and that $|\phi|$ is bounded by $M$. Denote $k_C = \lfloor \frac{\mathrm{e}^C}{\alpha} - \frac{1}{2} \rfloor$, the largest integer $k$ such that $\log((k + \frac{1}{2})\alpha) \leq C$. For small $\delta > 0$,

$$
\left| \int_{-\infty}^{\infty} \phi(a) \gamma \mu(\mathrm{d}a) - \sum_{k=1}^{\infty} \phi(x_k) \nu(x_k) \right|
$$

$$
\leq \left| \sum_{k=1}^{k_C} \int_{x_k + \delta}^{x_{k+1} - \delta} \phi(a) \gamma \mu(\mathrm{d}a) + \int_{-C}^{x_1 - \delta} \phi(a) \gamma \mu(\mathrm{d}a) \right|
$$

$$
+ \left| \sum_{k=1}^{k_C} \int_{x_k - \delta}^{x_k + \delta} \phi(a) \gamma \mu(\mathrm{d}a) - \sum_{k=1}^{k_C} \phi(x_k) \nu(x_k) \right|
$$

$$
\leq M \sum_{k=1}^{k_C} (\gamma \mu[x_k + \delta, x_{k+1} - \delta] + \gamma \mu[-C, x_1 - \delta])
$$

$$
+ \left| \sum_{k=1}^{k_C} \int_{x_k - \delta}^{x_k + \delta} \phi(a) \gamma \mu(\mathrm{d}a) - \sum_{k=1}^{k_C} \int_{x_k - \delta}^{x_k + \delta} \phi(x_k) \gamma \mu(\mathrm{d}a) \right|
$$

$$
+ \left| \sum_{k=1}^{k_C} \int_{x_k - \delta}^{x_k + \delta} \phi(x_k) \gamma \mu(\mathrm{d}a) - \sum_{k=1}^{k_C} \phi(x_k) \nu(x_k) \right|.
$$

As $\gamma \to 0$, the first term above goes to zero by (33) and (34); the second term can be arbitrarily small by letting $\delta$ be small; the third term goes to zero by (32). Hence,

$$\lim_{\gamma \to 0} \int_{-\infty}^{\infty} \phi(a)\gamma\mu(\mathrm{d}a) = \frac{1}{2\sqrt{\pi}} \sum_{k=1}^{\infty} \phi\left(\log\left(\left(k+\frac{1}{2}\right)\alpha\right)\right)\left(\log\frac{k+1}{k}\right)^2.$$

In particular, for any $\omega \in \Omega$ such that $L_t^a$ is continuous in $a$, $L_t^a$ is a continuous function of $a$ that has compact support. Therefore, by (27), almost surely,

$$\lim_{\gamma \to 0} \gamma\langle f(X), f(X)\rangle_T = \lim_{\gamma \to 0} \gamma\int_{-\infty}^{\infty} L_T^a\mu(\mathrm{d}a) = \frac{1}{2\sqrt{\pi}} \sum_{k=1}^{\infty} L_T^{\log((k+1/2)\alpha)}\left(\log\frac{k+1}{k}\right)^2. \qquad \square$$

**Proof of Theorem 3.** We borrow the notations from Jacod [17]: For $t_i = \frac{iT}{n}, i = 0, 1, \ldots, n$,

$$\xi_i^n := X_{t_i} - X_{t_{i-1}}; \qquad \chi_i^n := (f(X_{t_i}) - f(X_{t_{i-1}}))^2, \qquad \text{where } f(x) \text{ is defined in (28)},$$

$$R_n^k := \{(x,y) : f(x) = \log(k\alpha), f(y) = \log((k+1)\alpha) \text{ or}$$
$$f(x) = \log((k+1)\alpha), f(y) = \log(k\alpha)\},$$

$$R_n := \bigcup_{k=1}^{\infty} R_n^k, \qquad S_n := \mathbb{R}^2 \setminus R_n, \qquad T(a) := \{(x,y) : x < a \le y \text{ or } y < a \le x\},$$

$$T_n := \bigcup_{k=1}^{\infty} T(\log((k+1/2)\alpha)), \qquad \hat{W}_n := \frac{1}{\sqrt{n}}\sum_{i=1}^{n} I_{R_n}(X_{t_{i-1}}, X_{t_i}),$$

$$W_n := \sum_{i=1}^{n} \frac{1}{\sqrt{n}}\chi_i^n I_{S_n}(X_{t_{i-1}}, X_{t_i}).$$

If $(x,y) \in S_n$, then either $f(x) = f(y)$, or

$$|\exp(f(x)) - \exp(f(y))| > \alpha. \tag{35}$$

In the case of (35), without lost of generality, we can assume $\exp(x) > \exp(y) \ge \alpha$. Then

$$|f(x) - f(y)| = \log\frac{(\exp(x))^{(\alpha)}}{(\exp(y))^{(\alpha)}} \le \log\frac{\exp(x) + \alpha/2}{\exp(y) - \alpha/2}$$

$$\le \log\frac{\exp(x) + \exp(x)/2}{\exp(y) - \exp(y)/2} = \log 3 + |x - y|. \tag{36}$$

If, in addition, we have that both $\exp(x)$ and $\exp(y)$ are bounded by $M > 0$, then there exists $\theta$, $(x \wedge y) \le \theta \le (x \vee y)$ such that $\alpha \le |\exp(x) - \exp(y)| = \exp(\theta)|x - y| \le M|x - y|$. This implies that $|x - y| \ge \frac{\alpha}{M}$; hence, $\log 3 \le \frac{M\log 3}{\alpha}|x - y|$. By (36),

$$|f(x) - f(y)| \le \left(\frac{M\log 3}{\alpha} + 1\right)|x - y|. \tag{37}$$

It is easy to see that (37) holds for all $(x, y) \in S_n$ such that $\exp(x)$ and $\exp(y)$ are bounded by $M$. Therefore,

$$
\begin{aligned}
\mathrm{E}(W_n I_{\{\tau_{\log M} > T\}}) &= \sum_{i=1}^{n} \frac{1}{\sqrt{n}} E(\chi_i^n I_{S_n}(X_{t_{i-1}}, X_{t_i}) I_{\{\tau_{\log M} > T\}}) \\
&\leq \sum_{i=1}^{n} \frac{1}{\sqrt{n}} \left( \frac{M \log 3}{\alpha} + 1 \right)^2 \frac{\sigma^2 T}{n},
\end{aligned}
$$

which implies (also by making use of (16))

$$
W_n \to_P 0 \qquad \text{as } n \to \infty. \tag{38}
$$

On the other hand, for $i = 1, 2, \ldots, n$,

$$
\begin{aligned}
\mathrm{E}(I_{\{|\exp(X_{t_i}) - \exp(X_{t_{i-1}})| \geq \alpha\}}) &\leq \mathrm{E}(\exp(X_{(iT/n)}) - \exp(X_{((i-1)T/n)}))^2 / \alpha^2 \\
&= (\mathrm{E} \exp(2\sigma W_{iT/n}) + \mathrm{E} \exp(2\sigma W_{((i-1)T)/n}) \\
&\quad - 2\mathrm{E} \exp(2\sigma W_{((i-1)T)/n}) \mathrm{E} \exp(\sigma^2 W_{T/n})) / \alpha^2 \\
&= \frac{1}{\alpha^2} \exp\left( \frac{2\sigma^2 (i-1)T}{n} \right) \left( \exp\left( \frac{2\sigma^2 T}{n} \right) + 1 - 2\exp\left( \frac{\sigma^2 T}{2n} \right) \right) \\
&\leq \frac{1}{\alpha^2} \exp(2\sigma^2 T) \sum_{k=1}^{\infty} \frac{(2\sigma^2 T)^k - 2(\sigma^2 T/2)^k}{(k! n^k)} \\
&\leq \frac{1}{n} \left( \frac{\exp(2\sigma^2 T)(\exp(2\sigma^2 T) - \exp(\sigma^2 T/2))}{\alpha^2} \right),
\end{aligned}
$$

which implies

$$
\frac{1}{\sqrt{n}} \sum_{i=1}^{n} I_{\{|\exp(X_{t_i}) - \exp(X_{t_{i-1}})| \geq \alpha\}} \to_P 0 \qquad \text{as } n \to \infty. \tag{39}
$$

Note also that if $(X_{t_{i-1}}, X_{t_i}) \in R_n^k$, then $\chi_i^n = (\log \frac{k+1}{k})^2$. We have, for $k_l = \lfloor \frac{e^l}{\alpha} - \frac{1}{2} \rfloor$ (the greatest integer $k$ such that $\log((k + \frac{1}{2})\alpha) \leq l$),

$$
\begin{aligned}
\frac{1}{\sqrt{n}} [Y, Y]^{(\mathrm{all})} I_{\{\tau_l > T\}} &= \left( \hat{W}_n \sum_{k=1}^{\infty} I_{R_n^k}(X_{t_{i-1}}, X_{t_i}) \chi_i^n + W_n \right) I_{\{\tau_l > T\}} \\
&= \left( \sum_{k=1}^{k_l} (\hat{W}_n I_{R_n^k}(X_{t_{i-1}}, X_{t_i})) \cdot \left( \log \frac{k+1}{k} \right)^2 + W_n \right) I_{\{\tau_l > T\}}. \tag{40}
\end{aligned}
$$

We have $R_n \subset T_n$ and $|\exp(x) - \exp(y)| \geq \alpha$ when $(x, y) \in T_n \setminus R_n$. By (38), (39) and (40), we see that, for $\hat{W}'_n = \frac{1}{\sqrt{n}} \sum_{i=1}^n I_{T_n}(X_{t_{i-1}}, X_{t_i})$,

$$\operatorname*{plim}_{n \to \infty} \frac{1}{\sqrt{n}} [Y, Y]^{(\mathrm{all})} I_{\{\tau_l > T\}}$$

$$= \operatorname*{plim}_{n \to \infty} \sum_{k=1}^{k_l} \hat{W}'_n I_{T(\log((k+1/2)\alpha))}(X_{t_{i-1}}, X_{t_i}) \left( \log \frac{k+1}{k} \right)^2 I_{\{\tau_l > T\}}$$

$$= \operatorname*{plim}_{n \to \infty} \sum_{k=1}^{k_l} \frac{1}{\sqrt{n}} \sum_{i=1}^n I_{T(\log((k+1/2)\alpha))}(X_{t_{i-1}}, X_{t_i}) \left( \log \frac{k+1}{k} \right)^2 I_{\{\tau_l > T\}}$$

$$= \sum_{k=1}^{k_l} \left[ \left( \log \frac{k+1}{k} \right)^2 \operatorname*{plim}_{n \to \infty} \frac{1}{\sqrt{n}} \sum_{i=1}^n I_{T(\log((k+1/2)\alpha))}(X_{t_{i-1}}, X_{t_i}) \right] I_{\{\tau_l > T\}},$$

and for $k = 1, 2, 3, \ldots$, by Azaïs [4] and Jacod [17],

$$\frac{1}{\sqrt{n}} \sum_{i=1}^n I_{T(\log((k+1/2)\alpha))}(X_{t_{i-1}}, X_{t_i}) \xrightarrow{\mathbb{L}^2} \frac{1}{\sigma\sqrt{T}} \sqrt{\frac{2}{\pi}} L_T^{\log((k+1/2)\alpha)}.$$

Note that $l$ can be chosen to be arbitrarily large in the above argument. Therefore, by (16), we have

$$\operatorname*{plim}_{n \to \infty} \frac{1}{\sqrt{n}} [Y, Y]^{(\mathrm{all})} = \frac{1}{\sigma\sqrt{T}} \sqrt{\frac{2}{\pi}} \sum_{k=1}^{\infty} L_T^{\log((k+1/2)\alpha)} \left( \log \frac{k+1}{k} \right)^2 \qquad (41)$$

and

$$\operatorname*{plim}_{\bar{n} \to \infty} \frac{1}{\sqrt{\bar{n}}} [Y, Y]^{(\mathrm{avg})} = \frac{1}{\sigma\sqrt{T}} \sqrt{\frac{2}{\pi}} \sum_{k=1}^{\infty} L_T^{\log((k+1/2)\alpha)} \left( \log \frac{k+1}{k} \right)^2.$$

Applying these results to the TSRV (6), noting that (41) implies $\frac{\sqrt{\bar{n}}}{n} [Y, Y]^{(\mathrm{all})} \to_P 0$ by assumption (7), yields

$$\frac{1}{\sqrt{\bar{n}}} \widehat{\langle X, X \rangle}_T = \frac{1}{\sqrt{\bar{n}}} \left( [Y, Y]_T^{(\mathrm{avg})} - \frac{\bar{n}}{n} [Y, Y]_T^{(\mathrm{all})} \right)$$

$$\xrightarrow{P} \frac{1}{\sigma\sqrt{T}} \sqrt{\frac{2}{\pi}} \sum_{k=1}^{\infty} L_T^{\log((k+1/2)\alpha)} \left( \log \frac{k+1}{k} \right)^2. \qquad \square$$

## 4. Conclusion

We have shown in this paper that the robustness of the two scales realized volatility (TSRV) depends crucially on the deterministic part of the distortion through the function $f$ defined in (4). On the other hand, in terms of consistency and order of convergence, the TSRV is always robust to the random part of the error $(Y - f(X))$. In Section 3, we have studied a particular model of contamination, involving random error followed by rounding, and we have seen that, in this case, depending on parameters, the non-random distortion can be benign or problematic.

A lesson from our study is that there are really two candidates for the term 'volatility,' namely $\langle X, X \rangle_T$ and $\langle f(X), f(X) \rangle_T$, and that in some cases they can diverge substantially. Further investigation of what quantity one wishes to estimate necessitates more research into the use of realized volatility estimates in such applications as portfolio management, options trading and forecasting.

# Acknowledgements

We gratefully acknowledge the support of the National Science Foundation under grants DMS-0204639 and DMS-0604758. We would like to thank the editor and referees for their helpful remarks and suggestions that improved the paper.